\documentstyle[12pt]{article}
\def\simlt{\stackrel{<}{{}_\sim}}

\def\Eq{\begin{equation}}
\def\End{\end{equation}}

\def\Eqa{\begin{eqnarray}}
\def\Enda{\end{eqnarray}}

\def\Endl#1{\label{#1} \End}

\def\ord#1{{\cal O}(#1)}

\begin{document}
\renewenvironment{thebibliography}[1]
  { \begin{list}{\arabic{enumi}.}
    {\usecounter{enumi} \setlength{\parsep}{0pt}
     \setlength{\itemsep}{3pt} \settowidth{\labelwidth}{#1.}
     \sloppy
    }}{\end{list}}
\begin{titlepage}

\title{{\bf Calculating the bound on the
light Higgs mass in general SUSY models}\thanks{Talk given at the
XVI Kazimierz Meeting on Elementary Particles: {\em New Physics with
New Experiments}, 24-28 May 1993, Kazimierz (Poland).}}

\author{{\bf J. R. Espinosa} \thanks{Supported by a grant of Comunidad de
Madrid, Spain.}\\
Instituto de Estructura de la Materia, CSIC\\
Serrano 123, 28006-Madrid, Spain}

\date{}
\maketitle
\vspace{1.5cm}
\def\baselinestretch{1.15}
\begin{abstract}
\noindent
In the Minimal Supersymmetric Standard Model (MSSM) the existence
of an upper bound on the mass
of the $CP=+1$ lightest Higgs boson,
equal to $m_Z$ at tree--level and $\simlt 120\ GeV$ after the
inclusion of radiative corrections,
has important phenomenological consequences for Higgs searches.
A similar bound, independent of mass parameters other than
the electroweak scale, can be calculated
in supersymmetric models with an extended Higgs sector.
In models with arbitrary Higgs sectors
perturbative up to $10^{16}\ GeV$, we find,
including radiative corrections,
$m_h \simlt 155\ GeV$ for $m_t \simlt 190\ GeV$.
\end{abstract}

\thispagestyle{empty}

\vskip-15.5cm
\rightline{{\bf hep-ph/9309304}}
\rightline{{\bf IEM--FT--79/93}}
\rightline{{\bf September 1993}}
\vskip3in

\end{titlepage}

Supersymmetric models have more Higgs degrees of freedom than
the Standard Model but due to Supersymmetry their
Higgs sectors are also more
constrained $^1$. In particular, the MSSM
needs two Higgs doublets $H_1$, $H_2$ with opposite hypercharges
and from those, two scalar, one
pseudoscalar and two charged states show up in the Higgs spectrum.
{}From the fact that quartic couplings of the
Higgs bosons are given by the gauge couplings, the interesting tree--level
inequality results
\Eq
m_h\leq m_Z\mid \cos 2\beta\mid \ ,
\Endl{mhmssm}
where $m_h$ is the mass of the $CP=+1$ lightest Higgs scalar
$h^0$, $\tan\beta\equiv v_2/v_1$, and $v_i\equiv\langle H_i^o\rangle$.

The phenomenological consequences of the upper bound (\ref{mhmssm})
are clear: it would imply that LEP-200 should discover this Higgs $^2$.
However, it has been found recently that radiative corrections due to
top--stop loops can change (\ref{mhmssm}) sizeably.
These corrections can be calculated by different methods$^3$
and turn out to depend on the mass of the top $m_t$, $\tan\beta$ and
the scale of supersymmetry breaking $\Lambda_s$. This is shown in
Fig. 1a where the bound on $m_h$ includes two--loop radiative
corrections as calculated in Ref. 4.
For example, for $\Lambda_s \simlt 10\ TeV$ and
$m_t \simlt 150\ GeV$ one finds $m_h \simlt 115\ GeV$.

In a supersymmetric model with an enlarged Higgs sector, {\it i.e.}
with extra Higgs fields apart from the two doublets $H_1, H_2$ of the
MSSM, the bound (\ref{mhmssm}) changes but still exists.
It can be shown that this bound comes from the quartic couplings in
the Higgs potential and in the general case these are not only gauge
couplings (coming from D--terms as in the MSSM) but also Yukawa
couplings that can be present in the superpotential of the model (and
contribute to the potential through F--terms).

To be more specific, suppose that $H_1, H_2$ are the doublets that
take non--zero vevs ($\langle H_i^0\rangle = v_i^2$) and
couple to quarks and leptons. The
potential for their neutral components $H_1^0, H_2^0$ has some quartic
gauge couplings identical to the MSSM ones plus some new
quartic couplings from F--terms. The origin of these is clear: now
the superpotential can have trilinear terms of the form $\lambda\phi
H_i^0 H_j^0$ $(i,j=1,2)$ with $\phi$ some Higgs scalar.
For a superpotential that includes a part like
\Eq
f=\lambda_0 \phi_0 H_1^0 H_2^0
+\frac{1}{2}\lambda_1 \phi_1 H_1^0 H_1^0
+\frac{1}{2}\lambda_{-1} \phi_{-1} H_2^0 H_2^0 + \cdots ,
\Endl{sp}
the bound (\ref{mhmssm}) changes to$^5$
\Eq
m_h^2/v^2 \leq \frac{1}{2}(g^2+g'^2)\cos^2 2 \beta
+ \lambda_0^2 \sin^2  2\beta
+ \lambda_1^2 \cos^4 \beta
+\lambda_{-1}^2 \sin^4 \beta,
\Endl{bou}
where $v^2\equiv v_1^2+v_2^2$, and $g,\ g'$ are the $SU(2)\times
U(1)$ couplings. For $\lambda_{k}=0$ the bound
(\ref{mhmssm}) is recovered.

$SU(2)\times U(1)$ invariance requires the $\phi_k$'s to be singlets
or neutral components of $Y=0,\pm 1$ triplets. So that, concerning
this mass bound problem, the most general superpotential can be written as
\Eq
f_h=\vec{\lambda}_1 \cdot\vec{S}H_1 {\scriptstyle\circ} H_2+
\vec{\lambda}_2 \cdot \vec{\Sigma} H_2 {\scriptstyle\circ} H_1+\frac{1}{2}
\vec{\chi}_1 \cdot \vec{\Psi}_1 H_1 {\scriptstyle\circ} H_1+ \frac{1}{2}
\vec{\chi}_2 \cdot \vec{\Psi}_2 H_2 {\scriptstyle\circ} H_2+
\ord{\Sigma^3,...}.
\Endl{fhgen}
Here, we have gauge singlets $S^{(\sigma)}$, $\sigma=1,...,n_s$;
$SU(2)$ triplets $\Sigma^{(a)}$, $a=1,...,t_o$, with $Y=0$;
and $SU(2)$ triplets $\Psi^{(i)}_1$, $\Psi^{(i)}_2$,
$i=1,...,t_1$, with $Y=\pm 1$.
This superpotential gives a tree--level bound (\ref{bou}) of the
form$^7$
\Eq
m_h^2/v^2\leq\frac{1}{2}(g^2+g'^2)\cos^2 2\beta +
(\vec{\lambda}^2_{1}
+\frac{1}{2}\vec{\lambda}^2_{2})\sin^2 2
\beta+ \vec{\chi}^2_1\cos^4\beta+ \vec{\chi}^2_2\sin^4\beta.
\Endl{mhgen}
This bound is independent of supersymmetric mass terms or soft breaking
parameters. The gauge part is that of the MSSM and reproduces the
mass of the $Z$ provided that no other Higgs representations
contribute to it (if this is not the case this gauge contribution is
smaller and the bound stronger). The extra contribution involving the
Yukawa couplings is positive definite and so it rises the bound.
To obtain a numerical estimate of the latter these Yukawa couplings
must be bounded by triviality arguments. That is, requiring the theory to
remain perturbative up to some large scale
$\Lambda$ one
can compute the maximum allowed values of the Yukawas
at the electroweak scale. Inserting these values in Eq. (\ref{mhgen})
the tree-level upper bound (and the absolute one after inclusion
of radiative corrections) is obtained.

We can study the bound (\ref{mhgen}) in some cases of interest.

\noindent
{\bf {\it i)} MSSM + singlets.}
This is the simplest extension one can think of, possesing some
appealing theoretical properties, and having been
extensively studied in the literature$^{6-10}$.
We can choose the high scale $\Lambda$, up to which
the theory should remain perturbative, to be $\Lambda_{GUT}$ (the
inclusion of gauge singlets does not change the good unification
properties of the MSSM).
The bound after the inclusion of radiative corrections
is shown in Fig. 1b (we use the
renormalization group method$^{7,10}$; for alternative
calculations see Ref. 11.). One can see that
the bound is now weaker than that in the MSSM. In particular the absolute
upper bound is $145\ GeV$.

\noindent
{\bf {\it ii)} General model perturbative up to $\Lambda$.}
We can also address the problem of finding the absolute upper bound on
$m_h$ for supersymmetric models which remain perturbative below some fixed
scale $\Lambda$. To maximize (\ref{mhgen}) we will need to saturate
this scale, {\it i.e.} $\lambda^2(\Lambda)/4\pi\sim 1$, where
$\lambda$ is some generic coupling, and to slow the running of
the Yukawas that appear in (\ref{mhgen}) by choosing the right Higgs
representations ($n_s,\
t_{o,1}$). We can also add more doublets to saturate some gauge coupling and
this helps in slowing the running of the Yukawas. The radiative
corrections are included again using the RGE method and taking the scale
of SUSY breaking of order $1\ TeV$.
We performed this program$^7$ and obtained the absolute upper
bound for any model perturbative up to $\Lambda$. Fig. 1c shows this
bound for $\Lambda=10^{16}\ GeV$ (also studied in Ref. 12.).
The evolution of the bound with
$\Lambda$ can be seen in Fig. 1d where the MSSM case
and the model with singlets are also plotted for comparison.

\section*{Acknowledgements}
The content of this talk is based on work done in
collaboration with Mariano Quir\'os to whom
I want to express my deep gratitude.

\section*{Figure captions}
\begin{itemize}
\item[Fig.1:]
Bounds on the light Higgs mass in different SUSY models: {\it a)}
MSSM; {\it b)} MSSM + singlets; {\it c)} General model perturbative up to
$10^{16} \ GeV$; {\it d)} Models perturbative up to the scale $\Lambda$
indicated.
\end{itemize}

\end{document}